\documentclass[10pt,english]{llncs}
\usepackage[T1]{fontenc}
\usepackage[latin9]{inputenc}
\usepackage{array}
\usepackage{mathtools}
\usepackage{multirow}
\usepackage{amsmath}
\usepackage{amssymb}
\usepackage{graphicx}
\PassOptionsToPackage{normalem}{ulem}
\usepackage{ulem}

\makeatletter

\providecommand{\tabularnewline}{\\}

\usepackage{algorithm}
\usepackage{algpseudocode}

\usepackage{mathptm}

\makeatother

\usepackage{babel}
\begin{document}

\title{Proactive Defense Against Physical Denial of Service Attacks using
Poisson Signaling Games}

\author{Jeffrey Pawlick and Quanyan Zhu\thanks{This work is partially supported by an NSF IGERT grant through the
Center for Interdisciplinary Studies in Security and Privacy (CRISSP)
at New York University, by the grant CNS-1544782, EFRI-1441140, and
SES-1541164 from National Science Foundation (NSF) and DE-NE0008571
from the Department of Energy.}}

\institute{New York University Tandon School of Engineering\\Department of
Electrical and Computer Engineering\\6 MetroTech Center, Brooklyn,
NY 11201. \{jpawlick,quanyan.zhu\}@nyu.edu}

\titlerunning{Poisson Signaling Games for PDoS Attacks}
\maketitle
\begin{abstract}
While the Internet of things (IoT) promises to improve areas such
as energy efficiency, health care, and transportation, it is highly
vulnerable to cyberattacks. In particular, distributed denial-of-service
(DDoS) attacks overload the bandwidth of a server. But many IoT devices
form part of cyber-physical systems (CPS). Therefore, they can be
used to launch ``physical'' denial-of-service attacks (PDoS) in
which IoT devices overflow the ``physical bandwidth'' of a CPS.
In this paper, we quantify the population-based risk to a group of
IoT devices targeted by malware for a PDoS attack. In order to model
the recruitment of bots, we develop a ``Poisson signaling game,''
a signaling game with an unknown number of receivers, which have varying
abilities to detect deception. Then we use a version of this game
to analyze two mechanisms (legal and economic) to deter botnet recruitment.
Equilibrium results indicate that 1) defenders can bound botnet activity,
and 2) legislating a minimum level of security has only a limited
effect, while incentivizing active defense can decrease botnet activity
arbitrarily. This work provides a quantitative foundation for proactive
PDoS defense.
\end{abstract}

\section{Introduction to the IoT and PDoS Attacks\label{sec:Introduction}}

The Internet of things (IoT) is a ``dynamic global network infrastructure
with self-configuring capabilities based on standard and interoperable
communication protocols where physical and virtual `things' have
identities, physical attributes, and virtual personalities'' \cite{CERP2015}.
The IoT is 1) decentralized, 2) heterogeneous, and 3) connected to
the physical world. It is \emph{decentralized} because nodes have
``self-configuring capabilities,'' some amount of local intelligence,
and incentives which are not aligned with the other nodes. The IoT
is \emph{heterogeneous} because diverse ``things'' constantly enter
and leave the IoT, facilitated by ``standard and interoperable communication
protocols.'' Finally, IoT devices are \emph{connected to the physical
}world, \emph{i.e.}, they are part of cyber-physical systems (CPS).
For instance, they may influence behavior, control the flow of traffic,
and optimize home lighting. 
\begin{figure}
\begin{centering}
\includegraphics[width=0.55\columnwidth]{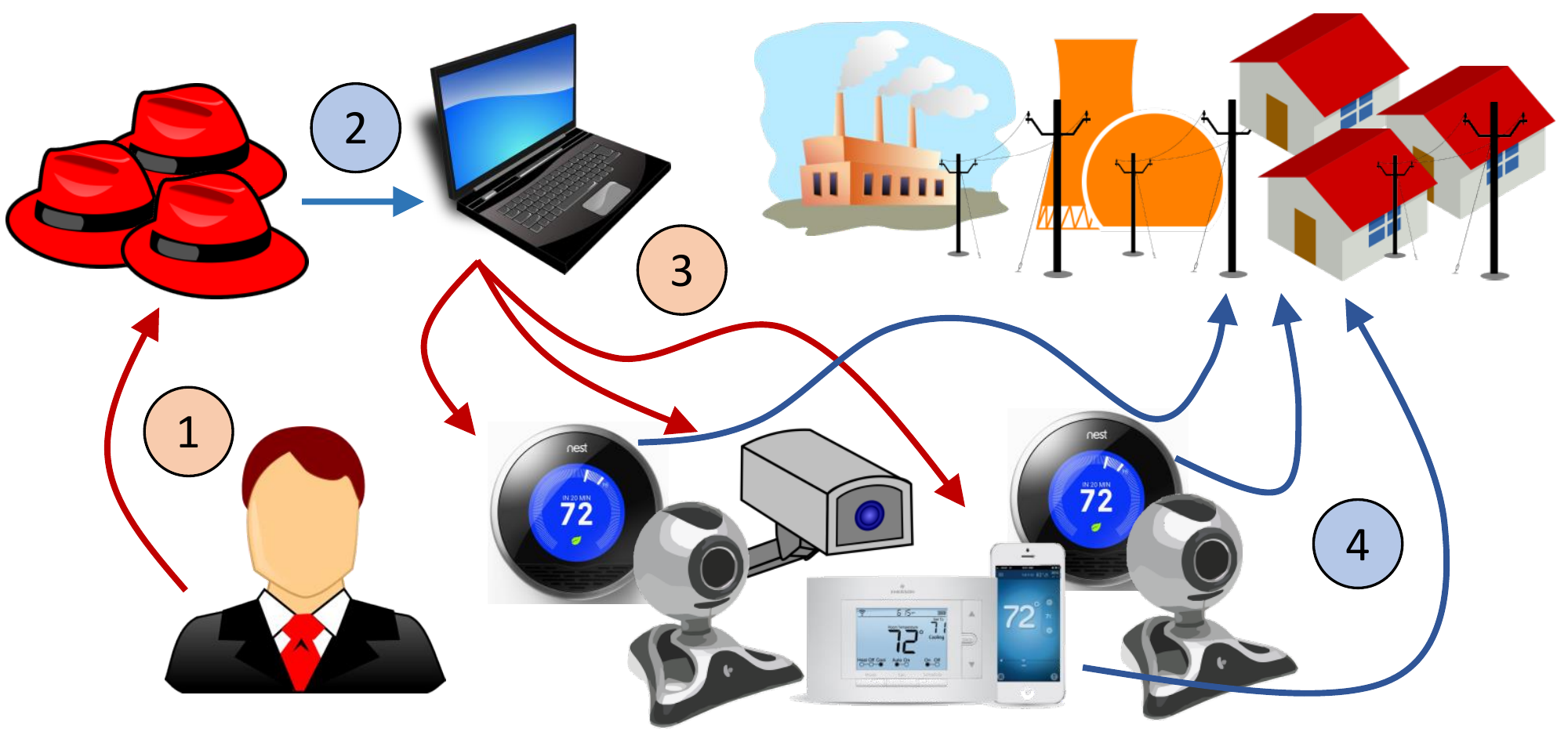} 
\par\end{centering}

\caption{\label{fig:DDoSsequence} Conceptual diagram of a PDoS attack. 1)
Attack sponsor hires botnet herder. 2) Botnet herder uses server to
manage recruitment. 3) Malware scans for vulnerable IoT devices and
begins cascading infection. 4) Botnet herder uses devices (\emph{e.g.,}
HVAC controllers) to deplete bandwidth of a cyber-physical service
(\emph{e.g.,} electrical power).}
\end{figure}

\subsection{Difficulties in Securing the Internet of Things}

While the IoT promises gains in efficiency, customization, and communication
ability, it also raises new challenges. One of these challenges is
security. The social aspect of IoT devices makes them vulnerable to
attack through social engineering. Moreover, the dynamic and heterogeneous
attributes of the IoT create a large attack surface. Once compromised,
these ``things'' serve as vectors for attack. The most notable example
has been the Mirai botnet attack on Dyn in 2016. Approximately 100,000
bots---largely belonging to the (IoT)---attacked the domain name server
(DNS) for Twitter, Reddit, Github, and the New York Times \cite{meyer2016dvr}.
A massive flow of traffic overwhelmed the bandwidth of the DNS.

\subsection{Denial of Cyber-Physical Service Attacks}

Since IoT devices are part of CPS, they also require physical ``bandwidth.''
As an example, consider the navigation app Waze\emph{ }\cite{wayz2017web}.
Waze uses real-time traffic information to find optimal navigation
routes. Due to its large number of users, the app also influences
traffic. If too many users are directed to one road, they can consume
the physical bandwidth of that road and cause unexpected congestion.
An attacker with insider access to Waze could use this mechanism to
manipulate transportation networks.

Another example can be found in healthcare. Smart lighting systems
(which deploy, \emph{e.g.}, \emph{time-of-flight} sensors) detect
falls of room occupants \cite{radke2016}. These systems alert emergency
responders about a medical situation in an assisted living center
or the home of someone who is aging. But an attacker could potentially
trigger many of these alerts at the same time, depleting the response
bandwidth of emergency personnel.

Such a threat could be called a denial of\emph{ cyber-physical }service
attack. To distinguish it from a cyber-layer DDoS, we also use the
acronym \emph{PDoS} (\emph{Physical} Denial of Service). Figure \ref{fig:DDoSsequence}
gives a conceptual diagram of a PDoS attack. In the rest of the paper,
we will consider one specific instance of a PDoS attack, although
our analysis is not limited to this example. We consider the infection
and manipulation of a population of IoT-based heating, ventilation,
and air conditioning (HVAC) controllers in order to cause a sudden
load shock to the power grid. Attackers either disable demand response
switches used for reducing peak load \cite{comverge2017demand}, or
they unexpectedly activate inactive loads. This imposes risks ranging
from frequency droop to load shedding and cascading failures. 
\begin{figure}
\begin{centering}
\includegraphics[width=0.6\columnwidth]{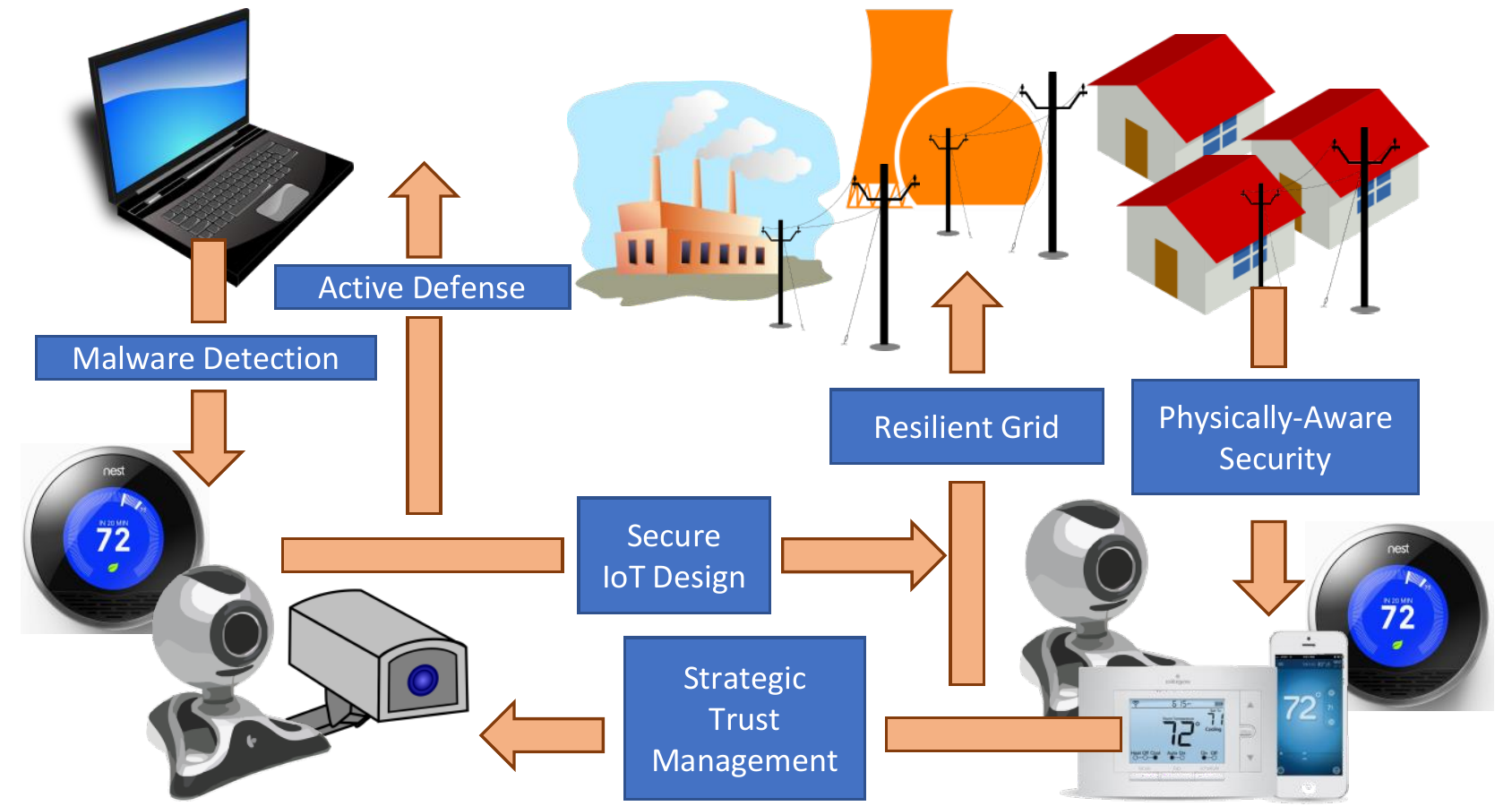}
\par\end{centering}

\caption{\label{fig:multiLayerDef}PDoS defense can be designed at multiple
layers. Malware detection and active defense can combat initial infection,
secure IoT design and strategic trust can reduce the spread of the
malware, and CPS can be resilient and physically-aware. We focus on
detection and active defense.}
\end{figure}

\subsection{Modeling the PDoS Recruitment Stage}

Defenses against PDoS can be designed at multiple layers (Fig. \ref{fig:multiLayerDef}).
The scope of this paper is limited to defense at the stage of botnet
recruitment, in which the attacker scans a wide range of IP addresses,
searching for devices with weak security settings. Mirai, for example,
does this by attempting logins with a dictionary of factory-default
usernames and passwords (\emph{e.g.} \texttt{root/admin, admin/admin,
root/123456}) \cite{herzberg2016breaking}. Devices in our mechanism
identify these suspicious login attempts and use active defense to
learn about the attacker or report his activity. 

In order to quantify the risk of malware infection, we combine two
game-theoretic models known as signaling games \cite{lewis2008convention,crawford1982strategic}
and Poisson games \cite{myerson1998population}. Signaling games model
interactions between two parties, one of which possesses information
unknown to the other party. While signaling games consider only two
players, we extend this model by allowing the number of target IoT
devices to be a random variable (r.v.) that follows a Poisson distribution.
This captures the fact that the malware scans a large number of targets.
Moreover, we allow the targets to have heterogeneous abilities to
detect malicious login attempts.

\subsection{Contributions and Related Work}

We make the following principle contributions: 
\begin{enumerate}
\item We describe an IoT attack called a \emph{denial of cyber-physical
service} (\emph{PDoS}).
\item We develop a general model called \emph{Poisson signaling games} (\emph{PSG})
which quantifies one-to-many signaling interactions.
\item We find the pure strategy equilibria of a version of the PSG model
for PDoS.
\item We analyze legal and economic mechanisms to deter botnet recruitment,
and find that 1) defenders can bound botnet activity, and 2) legislating
a minimum level of security has only a limited effect, while incentivizing
active defense, in principle, can decrease botnet activity arbitrarily. 
\end{enumerate}
Signaling games are often used to model deception and trust in cybersecurity
\cite{moghaddam2016game,pawlick2015flip,pawlickTOAPPEARdeception}.
Poisson games have also been used to model malware epidemics in large
populations \cite{hayel2017epidemic}. Wu \emph{et} \emph{al}. use
game theory to design defense mechanisms against DDoS attacks \cite{wu2010modeling}.
But the defense mechanisms mitigate the actual the flood of traffic
against a target system, while we focus on botnet recruitment. Bensoussan
\emph{et} \emph{al}. use a susceptible-infected-susceptible (SIS)
model to study the growth of a botnet \cite{bensoussan2010game}.
But IoT devices in our model maintain beliefs about the reliability
of incoming messages. In this way, our paper considers the need to
trust legitimate messages. Finally, \emph{load altering attacks} \cite{mohsenian2011distributed,amini2015dynamic}
to the power grid are an example of PDoS attacks. But PDoS attacks
can also deal with other resources.

In Section \ref{sec:GameModels}, we review signaling games and Poisson
games. In Section \ref{sec:Poisson-Signaling-Games}, we combine them
to create Poisson signaling games (PSG). In Section \ref{sec:Application-to-PDoS},
we apply PSG to quantify the population risk due to PDoS attacks.
Section \ref{sec:Equilibrium-Analysis} obtains the perfect Bayesian
Nash equilibria of the model. Some of these equilibria are harmful
for power companies and IoT users. Therefore, we design proactive
mechanisms to improve the equilibria in Section \ref{sec:Mechanism-Design}.
We underline the key contributions in Section \ref{sec:Discussion-of-Results}.

\section{Signaling Games and Poisson Games\label{sec:GameModels}}

This section reviews two game-theoretic models: signaling games and
Poisson games. In Section \ref{sec:Poisson-Signaling-Games}, we combine
them to create PSG. PSG can be used to model many one-to-many signaling
interactions in addition to PDoS.

\subsection{Signaling Games with Evidence\label{sub:Signaling-Games-with}}

Signaling games are a class of dynamic, two-player, information-asymmetric
games between a sender $S$ and a receiver $R$ (\emph{c.f.} \cite{lewis2008convention,crawford1982strategic}).
Signaling games \emph{with evidence} extend the typical definition
by giving receivers some exogenous ability to detect deception\footnote{This is based on the idea that deceptive senders have a harder time
communicating some messages than truthful senders. In interpersonal
deception, for instance, lying requires high cognitive load, which
may manifest itself in external gestures \cite{vrij2008increasing}. } \cite{pawlick2015deception}. They are characterized by the tuple
\[
\Phi_{\mathbf{SG}}=\left(X,M,E,A,q^{S},\delta,u^{S},u^{R}\right).
\]

First, $S$ posses some private information unknown to $R.$ This
private information is called a \emph{type}. The type could represent,
\emph{e.g.}, a preference, a technological capability, or a malicious
intent. Let the finite set $X$ denote the set of possible types,
and let $x\in X$ denote one particular type. Each type occurs with
a probability $q^{S}(x),$ where $q^{S}:\,X\to[0,1]$ such that (s.t.)
$\underset{x\in X}{\sum}q^{S}\left(x\right)=1$ and $\forall x\in X,\,q^{S}(x)\geq0.$ 

Based on his private information, $S$ communicates a \emph{message}
to the receiver. The message could be, \emph{e.g.}, a pull request,
the presentation of a certificate, or the execution of an action which
partly reveals the type. Let the finite set $M$ denote the set of
possible messages, and let $m\in M$ denote one particular type. In
general, $S$ can use a strategy in which he chooses various $m$
with different probabilities. We will introduce notation for these
\emph{mixed strategies} later.

In typical signaling games (\emph{e.g.} Lewis signaling games \cite{lewis2008convention,crawford1982strategic}
and signaling games discussed by Crawford and Sobel \cite{crawford1982strategic}),
$R$ only knows about $x$ through $m.$ But this suggests that deception
is undetectable. Instead, signaling games with evidence include a
\emph{detector}\footnote{This could literally be a hardware or software detector, such as email
filters which attempt to tag phishing emails. But it could also be
an an abstract notion meant to signify the innate ability of a person
to recognize deception.} which emits evidence $e\in E$ about the sender's type \cite{pawlick2015deception}.
Let $\delta:\,E\to[0,1]$ s.t. for all $x\in X$ and $m\in M,$ we
have $\underset{e\in E}{\sum}\delta(e\,|\,x,m)=1$ and $\delta(e\,|\,x,m)\geq0.$
Then $\delta(e\,|\,x,m)$ gives the probability with which the detector
emits evidence $e$ given type $x$ and message $m.$ This probability
is fixed, not a decision variable. Finally $A$ be a finite set of
\emph{actions}. Based on $m$ and $e,$ $R$ chooses some $a\in A.$
For instance, $R$ may choose to accept or reject a request represented
by the message. These can also be chosen using a mixed-strategy. 

In general, $x,$ $m,$ and $a$ can impact the utility of $S$ and
$R.$ Therefore, let $u^{S}:\,M\times A\to\mathbb{R}^{\left|X\right|}$
be a vector-valued function such that $u^{S}\left(m,a\right)=\left[u_{x}^{S}\left(m,a\right)\right]_{x\in X}.$
This is a column vector with entries $u_{x}^{S}(m,a).$ These entries
give the utility that $S$ of each receiver of type $x\in X$ obtains
for sending a message $m$ when the receiver plays action $a.$ Next,
define the utility function for $R$ by $u^{R}:\,X\times M\times A\to\mathbb{R},$
such that $u^{R}(x,m,a)$ gives the utility that $R$ receives when
a sender of type $x$ transmits message $m$ and $R$ plays action
$a.$

\subsection{Poisson Games\label{sub:Poisson-Games}}

Poisson games were introduced by Roger Myerson in 1998 \cite{myerson1998population}.
This class of games models interactions between an unknown number
of players, each of which belongs to one type in a finite set of types.
Modeling the population uncertainty using a Poisson r.v. is convenient
because merging or splitting Poisson r.v. results in r.v. which also
follow Poisson distributions. 

Section \ref{sec:Poisson-Signaling-Games} will combine signaling
games with Poisson games by considering a sender which issues a command
to a pool of an unknown number of receivers, which all respond at
once. Therefore, let us call the players of the Poisson game ``receivers,''
although this is not the nomenclature used in the original game. Poisson
games are characterized by the tuple
\[
\Phi^{\mathbf{PG}}=\left(\lambda,Y,q^{R},A,\tilde{u}^{R}\right).
\]

First, the population parameter $\lambda>0$ gives the mean and variance
of the Poisson distribution. For example, $\lambda$ may represent
the expected number of mobile phone users within range of a base station.
Let the finite set $Y$ denote the possible types of each receiver,
and let $y\in Y$ denote one of these types. Each receiver has type
$y$ with probability $q^{R}(y),$ where $\underset{y\in Y}{\sum}q^{R}(y)=1$
and $\forall y\in Y,\,q^{R}(y)>0.$ 

Because of the decomposition property of the Poisson r.v., the number
of receivers of each type $y\in Y$ also follows a Poisson distribution.
Based on her type, each receiver chooses an action $a$ in the finite
set $A.$ We have deliberately used the same notation as the action
for the signaling game, because these two actions will coincide in
the combined model. 

Utility functions in Poisson games are defined as follows. For $a\in A,$
let $c_{a}\in\mathbb{Z}_{+}$ (the set of non-negative integers) denote
the count of receivers which play action $a.$ Then let $c$ be a
column vector which contains entries $c_{a}$ for each $a\in A.$
Then $c$ falls within the set $\mathbb{Z}(A),$ the set of all possible
integer counts of the number of players which take each action. 

Poisson games assume that all receivers of the same type receive the
same utility. Therefore, let $\tilde{u}^{R}:\,A\times\mathbb{Z}(C)\to\mathbb{R}^{|Y|}$
be a vector-valued function such that $\tilde{u}^{R}\left(a,c\right)=\left[\tilde{u}_{y}^{R}\left(a,c\right)\right]_{y\in Y}.$
The entries $\tilde{u}_{y}^{R}\left(a,c\right)$ give the utility
that receivers of each type $y\in Y$ obtain for playing an action
$a$ while the vector of the total count of receivers that play each
action is given by $c.$ Note that this is different from the utility
function of receivers in the signaling game. Given the strategies
of the receivers, $c$ is also distributed according to a Poisson
r.v.

\section{Poisson Signaling Games\label{sec:Poisson-Signaling-Games}}

\begin{figure}
\begin{centering}
\includegraphics[width=0.45\columnwidth]{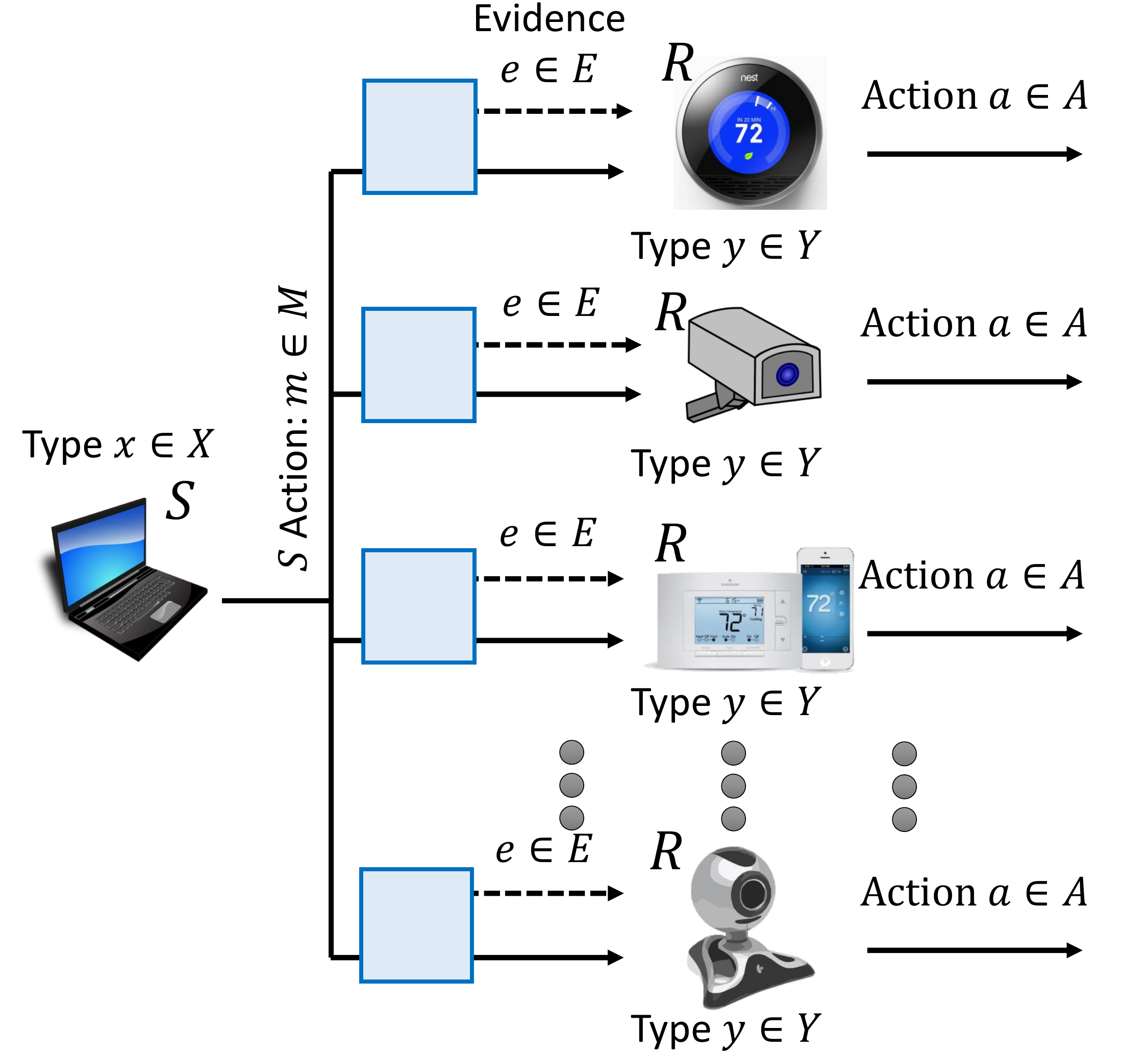} 
\par\end{centering}

\caption{\label{fig:actionFlow}PSG model the third stage of a PDoS attack.
A sender of type $x$ chooses an action $m$ which is observed by
an unknown number of receivers. The receivers have multiple types
$y\in Y.$ Each type may observe different evidence $e\in E.$ Based
on $m$ and $e,$ each type of receiver chooses an action $a.$}
\end{figure}

Figure \ref{fig:actionFlow} depicts Poisson signaling games (PSG).
PSG are characterized by combining $\Phi_{\mathbf{SG}}$ and $\Phi^{\mathbf{PG}}$
to obtain the tuple
\[
\Phi_{\mathbf{SG}}^{\mathbf{PG}}=\left(X,Y,M,E,A,\lambda,q,\delta,U^{S},U^{R}\right).
\]

\subsection{Types, Actions, and Evidence, and Utility\label{sub:Type-and-Action}}

As with signaling games and Poisson games, $X$ denotes the set of
types of $S,$ and $Y$ denotes the set of types of $R.$ $M,$ $E,$
and $A$ denote the set of messages, evidence, and actions, respectively.
The Poisson parameter is $\lambda.$ 

The remaining elements of $\Phi_{\mathbf{SG}}^{\mathbf{PG}}$ are
slightly modified from the signaling game or Poisson game. First,
$q:\,X\times Y\to[0,1]^{2}$ is a vector-valued function such that
$q\left(x,y\right)$ gives the probabilities $q^{S}(x),$ $x\in X,$
and $q^{R}(y),$ $y\in Y,$ of each type of sender and receiver, respectively. 

As in the signaling game, $\delta$ characterizes the quality of the
deception detector. But receivers differ in their ability to detect
deception. Various email clients, for example, may have different
abilities to identify phishing attempts. Therefore, in PSG, we define
the mapping by $\delta:\,E\to[0,1]^{|Y|},$ s.t. the vector $\delta\left(e\,|\,x,m\right)=\left[\delta_{y}\left(e\,|\,x,m\right)\right]_{y\in Y}$
gives the probabilities $\delta_{y}(e\,|\,x,m)$ with which each receiver
type $y$ observes evidence $e$ given sender type $x$ and message
$m.$ This allows each receiver type to observe evidence with different
likelihoods\footnote{In fact, although all receivers with the same type $y$ have the same
likelihood $\delta_{y}(e\,|\,x,m)$ of observing evidence $e$ given
sender type $x$ and message $m$, our formulation allows the receivers
to observe different actual realizations $e$ of the evidence.}. 

The utility functions $U^{S}$ and $U^{R}$ are also adjusted for
PSG. Let $U^{S}:\,M\times\mathbb{Z}(A)\to\mathbb{R}^{|X|}$ be a vector-valued
function s.t. the vector $U^{S}\left(m,c\right)=\left[U_{x}^{S}\left(m,c\right)\right]_{x\in X}$
gives the utility of senders of each type $x$ for sending message
$m$ if the count of receivers which choose each action is given by
$c.$ Similarly, let $U^{R}:\,X\times M\times A\times\mathbb{Z}(A)\to\mathbb{R}^{|Y|}$
be a vector-valued function s.t. $U^{R}\left(x,m,a,c\right)=\left[U_{y}^{R}\left(x,m,a,c\right)\right]_{y\in Y}$
gives the utility of receivers of each type $y\in Y.$ As earlier,
$x$ is the type of the sender, and $m$ is the message. But note
that $a$ denotes the action of \emph{this particular receiver}, while
$c$ denotes the count of overall receivers which choose each action.

\subsection{Mixed-Strategies and Expected Utility\label{sub:Mixed-Strategies-and-Beliefs}}

Next, we define the nomenclature for mixed-strategies and expected
utility functions. For senders of each type $x\in X,$ let $\sigma_{x}^{S}:\,M\to[0,1]$
be a mixed strategy such that $\sigma_{x}^{S}(m)$ gives the probability
with which he plays each message $m\in M.$ For each $x\in X,$ let
$\Sigma_{x}^{S}$ denote the set of possible $\sigma_{x}^{S}.$ We
have 
\[
\Sigma_{x}^{R}=\left\{ \bar{\sigma}\,|\,\underset{m\in M}{\sum}\bar{\sigma}\left(m\right)=1\;\text{and}\;\forall m\in M,\,\bar{\sigma}\left(m\right)\ge0\right\} .
\]

For receivers of each type $y\in Y,$ let $\sigma_{y}^{R}:\,A\to[0,1]$
denote a mixed strategy such that $\sigma_{y}^{R}(a\,|\,m,e)$ gives
the probability with which she plays action $a$ after observing message
$m$ and action $e.$  For each $y\in Y,$ the function $\sigma_{y}^{R}$
belongs to the set
\[
\Sigma_{y}^{R}=\left\{ \bar{\sigma}\,|\,\underset{a\in A}{\sum}\bar{\sigma}\left(a\right)=1\;\text{and}\;\forall a\in A,\,\bar{\sigma}\left(a\right)\ge0\right\} .
\]
In order to choose her actions, $R$ forms a belief about the sender
type $x.$ Let $\mu_{y}^{R}(x\,|\,m,e)$ denote the likelihood with
which each $R$ of type $y$ who observes message $m$ and evidence
$e$ believes that $S$ has type $x.$ In equilibrium, we will require
this belief to be consistent with the strategy of $S.$

Now we define the expected utilities that $S$ and each $R$ receive
for playing mixed strategies. Denote the expected utility of a sender
of type $x\in X$ by $\bar{U}_{x}^{S}:\,\Sigma_{x}^{S}\times\Sigma^{R}\to\mathbb{R}.$
Notice that all receiver strategies must be taken into account. This
expected utility is given by
\[
\bar{U}_{x}^{S}(\sigma_{x}^{S},\sigma^{R})=\underset{m\in M}{\sum}\,\underset{c\in\mathbb{Z}\left(A\right)}{\sum}\,\sigma_{x}^{S}\left(m\right)\mathbb{P}\left\{ c\,|\,\sigma^{R},x,m\right\} U_{x}^{S}(m,c).
\]
Here, $\mathbb{P}\{c\,|\,\sigma^{R},x,m\}$ is the probability with
which the vector $c$ gives the count of receivers that play each
action. Myerson shows that, due to the aggregation and decomposition
properties of the Poisson r.v., the entries of $c$ are also Poisson
r.v. \cite{myerson1998population}. Therefore, $\mathbb{P}\{c\,|\,\sigma^{R},x,m\}$
is given by 
\begin{equation}
\mathbb{P}\left\{ c\,|\,\sigma^{R},x,m\right\} =\prod_{a\in A}e^{\lambda_{a}}\frac{\lambda_{a}^{c_{a}}}{c_{a}!},\;\lambda_{a}=\lambda\underset{y\in Y}{\sum}\,\underset{e\in E}{\sum}\,q^{R}\left(y\right)\delta_{y}\left(e\,|\,x,m\right)\sigma_{y}^{R}\left(a\,|\,m,e\right).\label{eq:probLambdaA}
\end{equation}

Next, denote the expected utility of each receiver of type $y\in Y$
by $\bar{U}_{y}^{R}:\,\Sigma_{y}^{R}\times\Sigma^{R}\to\mathbb{R}.$
Here, $\bar{U}_{y}^{R}(\theta,\sigma^{R}\,|\,m,e,\mu_{y}^{R})$ gives
the expected utility when this particular receiver plays mixed strategy
$\theta\in\Sigma_{y}^{R}$ and the population of all types of receivers
plays the mixed-strategy vector $\sigma^{R}.$ The expected utility
is given by
\begin{multline}
\bar{U}_{y}^{R}(\theta,\sigma^{R}\,|\,m,e,\mu_{y}^{R})=\underset{x\in X}{\sum}\,\underset{a\in A}{\sum}\,\underset{c\in\mathbb{Z}\left(A\right)}{\sum}\,\\
\mu_{y}^{R}\left(x\,|\,m,e\right)\theta\left(a\,|\,m,e\right)\mathbb{P}\left\{ c\,|\,\sigma^{R},x,m\right\} U_{y}^{R}(x,m,a,c),\label{eq:expUr}
\end{multline}
 where again $\mathbb{P}\{c\,|\,\sigma^{R},x,m\}$ is given by Eq.
(\ref{eq:probLambdaA}).

\subsection{Perfect Bayesian Nash Equilibrium\label{sub:Perfect-Bayesian-Nash}}

First, since PSG are dynamic, we use an equilibrium concept which
involves \emph{perfection}. Strategies at each information set of
the game must be optimal for the remaining subgame \cite{fudenberg1991game}.
Second, since PSG involve incomplete information, we use a \emph{Bayesian}
concept. Third, since each receiver chooses her action without knowing
the actions of the other receivers, the Poisson stage of the game
involves a \emph{fixed point}. All receivers choose strategies which
best respond to the optimal strategies of the other receivers. Perfect
Bayesian Nash equilibrium (PBNE) is the appropriate concept for games
with these criteria \cite{fudenberg1991game}. 

Consider the two chronological stages of PSG. The second stage takes
place among the receivers. This stage is played with a given $m,$
$e,$ and $\mu^{R}$ determined by the sender (and detector) in the
first stage of the game. When $m,$ $e,$ and $\mu^{R}$ are fixed,
the interaction between all receivers becomes a standard Poisson game.
Define $BR_{y}^{R}:\,\Sigma^{R}\to\mathcal{P}(\Sigma_{y}^{R})$ (where
$\mathcal{P}(\mathbb{S})$ denotes the power set of $\mathbb{S}$)
such that the best response of a receiver of type $y$ to a strategy
profile $\sigma^{R}$ of the other receivers is given by the strategy
or set of strategies 
\begin{equation}
BR_{y}^{R}\left(\sigma^{R}\,|\,m,e,\mu_{y}^{R}\right)\triangleq\underset{\theta\in\Sigma_{y}^{R}}{\arg\max}\:\bar{U}_{y}^{R}\left(\theta,\sigma^{R}\,|\,m,e,\mu_{y}^{R}\right).\label{eq:whatIsBrR}
\end{equation}

The first stage takes place between the sender and the set of receivers.
If we fix the set of receiver strategies $\sigma^{R},$ then the problem
of a sender of type $x\in X$ is to choose $\sigma_{x}^{S}$ to maximize
his expected utility given $\sigma^{R}.$ The last criteria is that
the receiver beliefs $\mu^{R}$ must be consistent with the sender
strategies according to Bayes' Law. Definition \ref{def:pbne} applies
PBNE to PSG.
\begin{definition}
\label{def:pbne}(PBNE) Strategy and belief profile $(\sigma^{S*},\sigma^{R*},\mu^{R})$
is a PBNE of a PSG if all of the following hold \cite{fudenberg1991game}:
\begin{equation}
\forall x\in X,\;\sigma_{x}^{S*}\in\underset{\sigma_{x}^{S}\in\Sigma_{x}^{S}}{\arg\max}\:\bar{U}_{x}^{S}(\sigma_{x}^{S},\sigma^{R*}),\label{eq:optS}
\end{equation}
\begin{equation}
\forall y\in Y,\,\forall m\in M,\,\forall e\in E,\;\sigma_{y}^{R*}\in BR_{y}^{R}\left(\sigma^{R*}\,|\,m,e,\mu_{y}^{R}\right),\label{eq:brR}
\end{equation}
\begin{equation}
\forall y\in Y,\,\forall m\in M,\,\forall e\in E,\;\mu_{y}^{R}\left(d\,|\,m,e\right)\in\frac{\delta_{y}\left(e\,|\,d,m\right)\sigma_{d}^{S}\left(m\right)q^{S}\left(d\right)}{\underset{\tilde{x}\in X}{\sum}\,\delta_{y}\left(e\,|\,\tilde{x},m\right)\sigma_{\tilde{x}}^{S}\left(m\right)q^{S}\left(\tilde{x}\right)},\label{eq:beliefNonT}
\end{equation}
if $\underset{\tilde{x}\in X}{\sum}\,\delta_{y}\left(e\,|\,\tilde{x},m\right)\sigma_{\tilde{x}}^{S}\left(m\right)q^{S}\left(\tilde{x}\right)>0,$
and $\mu_{y}^{R}\left(d\,|\,m,e\right)\in\left[0,1\right],$ otherwise.
We also always have $\mu_{y}^{R}\left(l\,|\,m,e\right)=1-\mu_{y}^{R}\left(d\,|\,m,e\right).$ 
\end{definition}

Equation (\ref{eq:optS}) requires the sender to choose an optimal
strategy given the strategies of the receivers. Based on the message
and evidence that each receiver observes, Eq. (\ref{eq:brR}) requires
each receiver to respond optimally to the profile of the strategies
of the other receivers. Equation (\ref{eq:beliefNonT}) uses Bayes'
law (when possible) to obtain the posterior beliefs $\mu^{R}$ using
the prior probabilities $q^{S},$ the sender strategies $\sigma^{S},$
and the characteristics $\delta_{y},$ $y\in Y$ of the detectors
\cite{pawlick2015deception}.

\section{Application of PSG to PDoS \label{sec:Application-to-PDoS}}

Section \ref{sec:Poisson-Signaling-Games} defined PSG in general,
without specifying the members of the type, message, evidence, or
action sets. In this section, we apply PSG to the recruitment stage
of PDoS attacks. Table \ref{tab:Application-PSG-to-PDoS} summarizes
the nomenclature. 

$S$ refers to the agent which attempts a login attempt, while $R$
refers to the device. Let the set of sender types be given by $X=\{l,d\},$
where $l$ represents a legitimate login attempt, while $d$ represents
a malicious attempt. Malicious $S$ attempt to login to many devices
through a wide IP scan. This number is drawn from a Poisson r.v. with
parameter $\lambda.$ Legitimate $S$ only attempt to login to one
device at a time. Let the receiver types be $Y=\{k,o,v\}.$ Type $k$
represents weak receivers which have no ability to detect deception
and do not use active defense. Type $o$ represents strong receivers
which can detect deception, but do not use active defense. Finally,
type $v$ represents active receivers which can both detect deception
and use active defense.
\begin{table}
\caption{\label{tab:Application-PSG-to-PDoS}Application of PSG to PDoS Recruitment}

\centering{}%
\begin{tabular}{|c|c|}
\hline 
Set & Elements\tabularnewline
\hline 
\hline 
Type $x\in X$ of $S$ & $l:$ legitimate, $d:$ malicious\tabularnewline
\hline 
Type $y\in Y$ of $R$  & $k:$ no detection; $o:$ detection; $v:$ detection \& active defense\tabularnewline
\hline 
Message $m\in M$ of $S$ & $m=\{m^{1},m^{2},\ldots\},$ a set of $|m|$ password strings\tabularnewline
\hline 
Evidence $e\in E$ & $b:$ suspicious, $n:$ not suspicious\tabularnewline
\hline 
Action $a\in A$ of $R$ & $t:$ trust, $g:$ lockout, $f:$ active defense\tabularnewline
\hline 
\end{tabular}
\end{table}

\subsection{Messages, Evidence Thresholds, and Actions}

Messages consist of sets of consecutive unsuccessful login attempts.
They are denoted by $m=\{m^{1},m^{2},\ldots\},$ where each $m^{1},m^{2},\ldots$
is a string entered as an attempted password\footnote{A second string can also be considered for the username.}.
For instance, botnets similar to Mirai choose a list of default passwords
such as \cite{herzberg2016breaking} 
\[
m=\left\{ \texttt{admin},\texttt{888888},\texttt{123456},\texttt{default},\texttt{support}\right\} .
\]
Of course, devices can lockout after a certain number of unsuccessful
login attempts. Microsoft Server 2012 recommends choosing a threshold
at $5$ to $9$ \cite{microsoft2014lockout}. Denote the lower end
of this range by $\tau_{L}=5.$ Let us allow all attempts with $|m|<\tau_{L}.$
In other words, if a user successfully logs in before $\tau_{L},$
then the PSG does not take place. (See Fig. \ref{fig:prunedTree}.)

The PSG takes place for $|m|\geq\tau_{L}.$ Let $\tau_{H}=9$ denote
the upper end of the Microsoft range. After $\tau_{L},$ $S$ may
persist with up to $\tau_{H}$ login attempts, or he may not persist.
Let $p$ denote persist, and $w$ denote not persist. Our goal is
to force malicious $S$ to play $w$ with high probability.

For $R$ of types $o$ and $v,$ if $S$ persists and does not successfully
log in with $|m|\leq\tau_{H}$ login attempts, then $e=b.$ This signifies
a suspicious login attempt. If $S$ persists and does successfully
login with $|m|\leq\tau_{H}$ attempts, then $e=n,$ \emph{i.e.},
the attempt is not suspicious\footnote{For strong and active receivers, $\delta_{y}\left(b\,|\,d,p\right)>\delta_{y}\left(b\,|\,l,p\right),$
$y\in\{o,v\}.$ That is, these receivers are more likely to observe
suspicious evidence if they are interacting with a malicious sender
than if they are interacting with a legitimate sender. Mathematically,
$\delta_{k}(b\,|\,d,p)=\delta_{k}(b\,|\,l,p)$ signifies that type
$k$ receivers do not implement a detector.}. 

If a user persists, then the device $R$ must choose an action $a.$
Let $a=t$ denote trusting the user, \emph{i.e.}, allowing login attempts
to continue. Let $a=g$ denote locking the device to future login
attempts. Finally, let $a=f$ denote using an active defense such
as reporting the suspicious login attempt to an Internet service provider
(ISP), recording the attempts in order to gather information about
the possible attacker, or attempting to block the offending IP address.

\subsection{Characteristics of PDoS Utility Functions\label{sub:Characteristics-of-PDoS}}

The nature of PDoS attacks implies several features of the utility
functions $U^{S}$ and $U^{R}.$ These are listed in Table \ref{tab:PDoS-Characteristics}.
Characteristic 1 (C1) states that if $S$ does not persist, both players
receive zero utility. C2 says that $R$ also receives zero utility
if $S$ persists and $R$ locks down future logins. Next, C3 states
that receivers of all types receive positive utility for trusting
a benign login attempt, but negative utility for trusting a malicious
login attempt. We have assumed that only type $v$ receivers use active
defense; this is captured by C4. Finally, C5 says that type $v$ receivers
obtain positive utility for using active defense against a malicious
login attempt, but negative utility for using active defense against
a legitimate login attempt. Clearly, C1-C5 are all natural characteristics
of PDoS recruitment.
\begin{table}
\caption{\label{tab:PDoS-Characteristics}Characteristics of PDoS Utility Functions }

\centering{}%
\begin{tabular}{|c|c|}
\hline 
\# & Notation\tabularnewline
\hline 
\hline 
\multirow{2}{*}{C1} & $\forall x\in X,\:y\in Y\:,a\in A,\:c\in\mathbb{Z}\left(A\right),$\tabularnewline
 & $U_{x}^{S}\left(w,c\right)=U_{y}^{R}(x,w,a,c)=0.$\tabularnewline
\hline 
\multirow{2}{*}{C2} & $\forall x\in X,\:y\in Y,\,c\in\mathbb{Z}\left(A\right),$\tabularnewline
 & $U_{y}^{R}(x,p,g,c)=0.$\tabularnewline
\hline 
\multirow{2}{*}{C3} & $\forall y\in Y,\:c\in\mathbb{Z}\left(A\right),$\tabularnewline
 & $U_{y}^{R}(d,p,t,c)<0<U_{y}^{R}(l,p,t,c).$\tabularnewline
\hline 
\multirow{2}{*}{C4} & $\forall x\in X,\,c\in\mathbb{Z}\left(A\right),$\tabularnewline
 & $U_{k}^{R}(x,p,f,c)=U_{o}^{R}(x,p,f,c)=-\infty.$\tabularnewline
\hline 
\multirow{2}{*}{C5} & $\forall c\in\mathbb{Z}\left(A\right),$\tabularnewline
 & $\;U_{v}^{R}(l,p,f,c)<0<U_{v}^{R}(d,p,f,c).$\tabularnewline
\hline 
\end{tabular}
\end{table}

\subsection{Modeling the Physical Impact of PDoS Attacks\label{sub:Modeling-Physical-Impact}}

The quantities $c_{t},$ $c_{g},$ and $c_{f}$ denote, respectively,
the number of devices that trust, lock down, and use active defense.
Define the function $Z:\,\mathbb{Z}(A)\to\mathbb{R}$ such that $Z(c)$
denotes the load shock that malicious $S$ cause based on the count
$c.$  $Z(c)$ is clearly non-decreasing in $c_{t},$ because each
device that trusts the malicious sender becomes infected and can impose
some load shock to the power grid. 
\begin{figure}
\begin{centering}
\includegraphics[width=0.45\columnwidth]{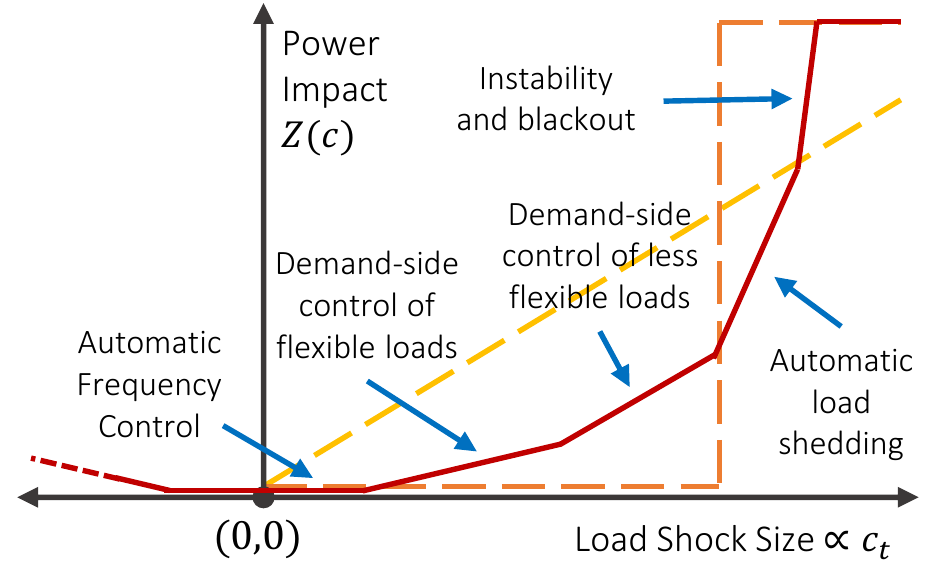}
\par\end{centering}

\caption{\label{fig:impactVshock}Conceptual relationship between load shock
size and damage to the power grid. Small shocks are mitigated through
automatic frequency control or demand-side control of flexible loads.
Large shocks can force load shedding or blackouts. }
\end{figure}

The red (solid) curve in Fig. \ref{fig:impactVshock} conceptually
represents the mapping from load shock size to damage caused to the
power grid based on the mechanisms available for regulation. Small
disturbances are regulated using automatic frequency control. Larger
disturbances can significantly decrease frequency and should be mitigated.
Grid operators have recently offered customers \emph{load control
switches}, which automatically deactivate appliances in response to
a threshold frequency decrease \cite{hammerstrom2007appliance}. But
the size of this voluntary demand-side control is limited. Eventually,
operators impose involuntary load shedding (\emph{i.e.}, rolling blackouts).
This causes higher inconvenience. In the worst case, transient instability
leads to cascading failures and blackout \cite{glover2012power}.

The yellow and orange dashed curves in Fig. \ref{fig:impactVshock}
provide two approximations to $Z(c)$. The yellow curve, $\tilde{Z}_{\text{lin}}(c),$
is linear in $c^{t}.$ We have $\tilde{Z}_{\text{lin}}(c)=\omega_{d}^{t}c^{t},$
where $\omega_{d}^{t}$ is a positive real number. The orange curve,
$\tilde{Z}_{\text{step}}(c),$ varies according to a step function,
\emph{i.e.}, $Z(c)=\Omega_{d}^{t}\mathbf{1}_{\{c_{t}>\tau_{t}\}},$
where $\Omega_{d}^{t}$ is a positive real number and $\mathbf{1}_{\{\bullet\}}$
is the indicator function. In this paper, we derive solutions for
the linear approximation. Under this approximation, the utility of
malicious $S$ is given by 
\[
U_{d}^{S}\left(m,c\right)=\tilde{Z}_{\text{lin}}(c)+\omega_{d}^{g}c_{g}+\omega_{d}^{f}c_{f}=\omega_{d}^{t}c_{t}+\omega_{d}^{g}c_{g}+\omega_{d}^{f}c_{f}.
\]
where $\omega_{d}^{g}<0$ and $\omega_{d}^{f}<0$ represent the utility
to malicious $S$ for each device that locks down or uses active defense,
respectively. 

Using $\tilde{Z}_{\text{lin}}(c),$ the decomposition property of
the Poisson r.v. simplifies $\bar{U}_{x}^{S}(\sigma_{x}^{S},\sigma^{R}).$
We show in Appendix \ref{apdx:simpSeUtil} that the sender's expected
utility depends on the expected values of each of the Poisson r.v.
that represent the number of receivers who choose each action $c_{a},$
$a\in A.$ The result is that 
\begin{equation}
\bar{U}_{x}^{S}(\sigma_{x}^{S},\sigma^{R})=\lambda\sigma_{x}^{S}\left(p\right)\underset{y\in Y}{\sum}\,\underset{e\in E}{\sum}\,\underset{a\in A}{\sum}\,q^{R}\left(y\right)\delta_{y}\left(e\,|\,x,p\right)\sigma_{y}^{R}\left(a\,|\,p,e\right)\omega_{x}^{a}.\label{eq:uBarStract}
\end{equation}

Next, assume that the utility of each receiver does not depend directly
on the actions of the other receivers. (In fact, the receivers are
still endogenously coupled through the action of $S.$) Abusing notation
slightly, we drop $c$ (the count of receiver actions) in $U_{y}^{R}(x,m,a,c)$
and $\sigma^{R}$ (the strategies of the other receivers) in $\bar{U}_{y}^{R}(\theta,\sigma^{R}\,|\,m,e,\mu_{y}^{R}).$
Equation (\ref{eq:expUr}) is now 
\[
\bar{U}_{y}^{R}\left(\theta\,|\,m,e,\mu_{y}^{R}\right)=\underset{x\in X}{\sum}\,\underset{a\in\left\{ t,f\right\} }{\sum}\,\mu_{y}^{R}\left(x\,|\,m,e\right)\theta\left(a\,|\,m,e\right)\mathcal{U}_{y}^{R}\left(x,m,a\right).
\]

\section{Equilibrium Analysis\label{sec:Equilibrium-Analysis}}

In this section, we obtain the equilibrium results by parameter region.
In order to simplify analysis, without loss of generality, let the
utility functions be the same for all receiver types (except when
$a=f$), \emph{i.e.}, $\forall x\in X,\;U_{k}^{R}(x,p,t)=U_{o}^{R}(x,p,t)=U_{v}^{R}(x,p,t).$
Also without loss of generality, let the quality of the detectors
for types $y\in\{o,v\}$ be the same: $\forall e\in E,\,x\in X,\;\delta_{o}(e\,|\,x,p)=\delta_{v}(e\,|\,x,p).$
\begin{figure}
\begin{centering}
\includegraphics[width=0.7\textwidth]{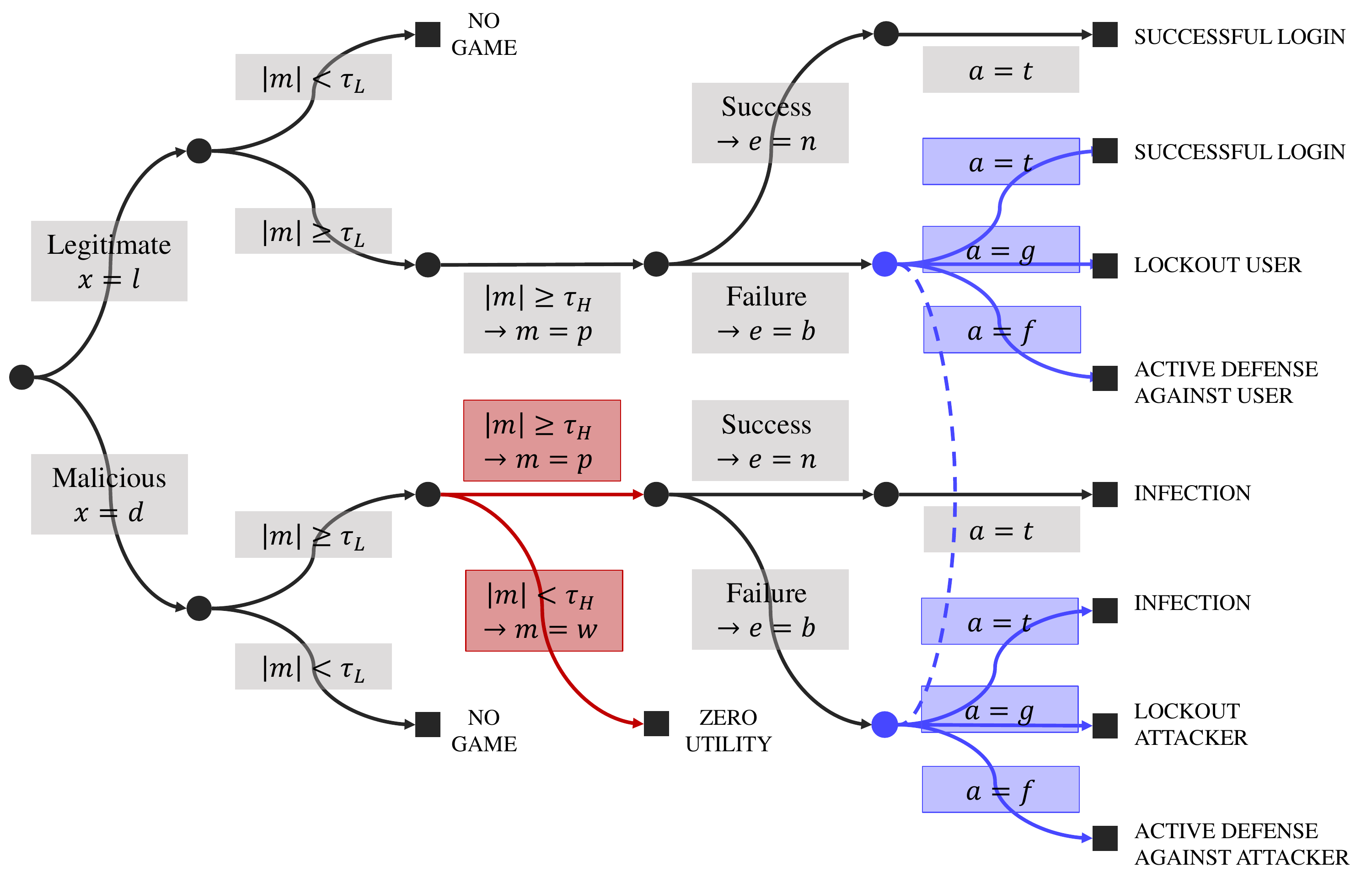} 
\par\end{centering}

\caption{\label{fig:prunedTree}Model of a PSG under Lemma \ref{lem:eqAfterAssump}.
Only one of many $R$ is depicted. After the types $x$ and $y,$
of $S$ and $R$, respectively, are drawn, $S$ chooses whether to
persist beyond $\tau_{L}$ attempts. Then $R$ chooses to trust, lockout,
or use active defense against $S$ based on whether $S$ is successful.
Lemma \ref{lem:eqAfterAssump} determines all equilibrium strategies
except $\sigma_{d}^{S*}\left(\bullet\right),$ $\sigma_{o}^{R*}\left(\bullet\,|\,p,b\right),$
and $\sigma_{v}^{R*}\left(\bullet\,|\,p,b\right),$ marked by the
blue and red items.}
\end{figure}

\subsection{PSG Parameter Regime}

We now obtain equilibria for a natural regime of the PSG parameters.
First, assume that legitimate senders always persist: $\sigma_{l}^{S}(p)=1.$
This is natural for our application, because IoT HVAC users will always
attempt to login. Second, assume that $R$ of all types trust login
attempts which appear to be legitimate (\emph{i.e.}, give evidence
$e=n$). This is satisfied for 
\begin{equation}
q^{S}\left(d\right)<\frac{U_{k}^{R}(l,p,t)}{U_{k}^{R}(l,p,t)-U_{k}^{R}(d,p,t)}.\label{eq:A2}
\end{equation}
Third, we consider the likely behavior of $R$ of type $o$ when a
login attempt is suspicious. Assume that she will lock down rather
than trust the login. This occurs under the parameter regime 
\begin{equation}
q^{S}\left(d\right)>\frac{\tilde{U}_{o}^{R}\left(l,p,t\right)}{\tilde{U}_{o}^{R}\left(l,p,t\right)-\tilde{U}_{o}^{R}\left(d,p,t\right)},\label{eq:A3}
\end{equation}
using the shorthand notation 
\[
\tilde{U}_{o}^{R}\left(l,p,t\right)=U_{o}^{R}\left(l,p,t\right)\delta_{0}\left(b\,|\,l,p\right),\;\;\tilde{U}_{o}^{R}\left(d,p,t\right)=U_{o}^{R}\left(d,p,t\right)\delta_{0}\left(b\,|\,d,p\right).
\]

The fourth assumption addresses the action of $R$ of type $v$ when
a login attempt is suspicious. The optimal action depends on her belief
$\mu_{o}^{R}(d\,|\,p,b)$ that $S$ is malicious. The belief, in turn,
depends on the mixed-strategy probability with which malicious $S$
persist.  We assume that there is some $\sigma_{d}^{S}(p)$ for which
$R$ should lock down ($a=g$). This is satisfied if there exists
a real number $\phi\in[0,1]$ such that, given\footnote{We abuse notation slightly to write $\bar{U}_{v}^{R}(a\,|\,m,e,\mu_{y}^{R})$
for the expected utility that $R$ of type $v$ obtains by playing
action $a.$ } $\sigma_{d}^{S}(p)=\phi,$
\begin{equation}
\bar{U}_{v}^{R}(t\,|\,p,b,\mu_{v}^{R})>0,\;\;\bar{U}_{v}^{R}(f\,|\,p,b,\mu_{v}^{R})>0.\label{eq:A4}
\end{equation}
 This simplifies analysis, but can be removed if necessary.

Lemma \ref{lem:eqAfterAssump} summarizes the equilibrium results
under these assumptions. Legitimate $S$ persist, and $R$ of type
$o$ lock down under suspicious login attempts. All receiver types
trust login attempts which appear legitimate. $R$ of type $k,$ since
she cannot differentiate between login attempts, trusts all of them.
The proof follows from the optimality conditions in Eq. (\ref{eq:optS}-\ref{eq:beliefNonT})
and the assumptions in Eq. (\ref{eq:A2}-\ref{eq:A4}). 
\begin{lemma}
\label{lem:eqAfterAssump}(Constant PBNE Strategies) If $\sigma_{d}^{S}(p)=1$
and Eq. (\ref{eq:A2}-\ref{eq:A4}) hold, then the following equilibrium
strategies are implied: 
\[
\sigma_{l}^{S*}\left(p\right)=1,\;\sigma_{o}^{R*}\left(g\,|\,p,b\right)=1,\;\sigma_{k}^{R*}\left(t\,|\,p,b\right)=1,
\]
\[
\sigma_{o}^{R*}\left(t\,|\,p,n\right)=\sigma_{v}^{R*}\left(t\,|\,p,n\right)=\sigma_{k}^{R*}\left(t\,|\,p,n\right)=1.
\]

\end{lemma}

Figure \ref{fig:prunedTree} depicts the results of Lemma \ref{lem:eqAfterAssump}.
The remaining equilibrium strategies to be obtained are denoted by
the red items for $S$ and the blue items for $R.$ These strategies
are $\sigma_{o}^{R*}(\bullet\,|\,p,b),$ $\sigma_{v}^{R*}(\bullet\,|\,p,b),$
and $\sigma_{d}^{S*}(p).$ Intuitively, $\sigma_{d}^{S*}(p)$ depends
on whether $R$ of type $o$ and type $v$ will lock down and/or use
active defense to oppose suspicious login attempts.

\subsection{Equilibrium Strategies}

The remaining equilibrium strategies fall into four parameter regions.
In order to delineate these regions, we define two quantities.

Let $TD_{v}^{R}(U_{v}^{R},\delta_{v})$ denote a threshold which determines
the optimal action of $R$ of type $v$ if $\sigma_{d}^{S}(p)=1.$
If $q^{S}(d)>TD_{v}^{R}(U_{v}^{R},\delta_{v}),$ then the receiver
uses active defense with some probability. Equation (\ref{eq:whatIsBrR})
can be used to show that 
\[
TD_{v}^{R}\left(U_{v}^{R},\delta_{v}\right)=\frac{\tilde{U}_{v}^{R}\left(l,p,f\right)}{\tilde{U}_{v}^{R}\left(l,p,f\right)-\tilde{U}_{v}^{R}\left(d,p,f\right)},
\]
where we have used the shorthand notation: 
\[
\tilde{U}_{v}^{R}\left(l,p,f\right)\coloneqq U_{v}^{R}\left(l,p,f\right)\delta_{v}\left(b\,|\,l,p\right),\;\;\tilde{U}_{v}^{R}\left(d,p,f\right)\coloneqq U_{v}^{R}\left(d,p,f\right)\delta_{v}\left(b\,|\,d,p\right).
\]

Next, let $BP_{d}^{S}\left(\omega_{d},q^{R},\delta\right)$ denote
the benefit which $S$ of type $d$ receives for choosing $m=p,$
\emph{i.e.}, for persisting. We have 
\[
BP_{d}^{S}\left(\omega_{d},q^{R},\delta\right)\coloneqq\underset{y\in Y}{\sum}\,\underset{e\in E}{\sum}\,\underset{a\in A}{\sum}q^{R}\left(y\right)\delta_{y}\left(e\,|\,d,p\right)\sigma_{y}^{R}\left(a\,|\,p,e\right)\omega_{d}^{a}.
\]
If this benefit is negative, then $S$ will not persist. Let $BP_{d}^{S}\left(\omega_{d},q^{R},\delta\,|\,a_{k},a_{o},a_{v}\right)$
denote the benefit of persisting when receivers use the pure strategies:
\[
\sigma_{k}^{R}\left(a_{k}\,|\,p,b\right)=\sigma_{o}^{R}\left(a_{o}\,|\,p,b\right)=\sigma_{v}^{R}\left(a_{v}\,|\,p,b\right)=1.
\]
We now have Theorem \ref{thm:eqInRegions}, which predicts the risk
of malware infection in the remaining parameter regions. The proof is in Appendix \ref{apdx:Thm1}. 
\begin{theorem}
\label{thm:eqInRegions}(PBNE within Regions) If $\sigma_{d}^{S}(p)=1$
and Eq. (\ref{eq:A2}-\ref{eq:A4}) hold, then $\sigma_{o}^{R*}(\bullet\,|\,p,b),$
$\sigma_{v}^{R*}(\bullet\,|\,p,b),$ and $\sigma_{d}^{S*}(p)$ vary
within the four regions listed in Table \ref{tab:eqRegion}.
\end{theorem}

In the \emph{status quo} equilibrium, strong and active receivers
lock down under suspicious login attempts. But this is not enough
to deter malicious senders from persisting. We call this the status
quo because it represents current scenarios in which botnets infect
vulnerable devices but incur little damage from being locked out of
secure devices. This is a poor equilibrium, because $\sigma_{d}^{S*}(p)=1.$ 

In\emph{ }the \emph{active deterrence} equilibrium, lockouts are not
sufficient to deter malicious $S$ from fully persisting. But since
$q^{S}(d)>TD_{v}^{R},$ $R$ of type $v$ use active defense. This
is enough to deter malicious $S:$ $\sigma_{d}^{S*}(p)<1.$ In this
equilibrium, $R$ of type $o$ always locks down: $\sigma_{o}^{R*}\left(g\,|\,p,b\right)=1.$
$R$ of type $v$ uses active defense with probability 
\begin{equation}
\sigma_{v}^{R*}\left(f\,|\,p,b\right)=\frac{\omega_{d}^{t}q^{R}\left(k\right)+\omega_{d}^{g}\left(q^{R}\left(o\right)+q^{R}\left(v\right)\right)}{\left(\omega_{d}^{g}-\omega_{d}^{f}\right)q^{R}\left(v\right)\delta_{v}\left(v\,|\,d,p\right)},\label{eq:eq4ProbF}
\end{equation}
and otherwise locks down: $\sigma_{v}^{R*}\left(g\,|\,p,b\right)=1-\sigma_{v}^{R*}\left(f\,|\,p,b\right).$
Deceptive $S$ persist with reduced probability 
\begin{equation}
\sigma_{d}^{S*}\left(p\right)=\frac{1}{q^{S}\left(d\right)}\left(\frac{\tilde{U}_{v}^{R}\left(l,p,f\right)}{\tilde{U}_{v}^{R}\left(l,p,f\right)-\tilde{U}_{v}^{R}\left(d,p,f\right)}\right).\label{eq:eq4decep}
\end{equation}

In the \emph{resistant attacker} equilibrium, $q^{S}(d)>TD_{v}^{R}.$
Therefore, $R$ of type $v$ use active defense. But $BP_{d}^{S}(\bullet\,|\,t,g,f)>0,$
which means that the active defense is not enough to deter malicious
senders. This is a ``hopeless'' situation for defenders, since all
available means are not able to deter malicious senders. We still
have $\sigma_{d}^{S*}(p)=1.$

In the \emph{vulnerable attacker} equilibrium, there is no active
defense. But $R$ of type $o$ and type $v$ lock down under suspicious
login attempts, and this is enough to deter malicious $S,$ because
$BP_{d}^{S}(\bullet\,|\,t,g,g)<0.$ $R$ of types $o$ and $v$ lock
down with probability
\begin{equation}
\sigma_{o}^{R*}\left(g\,|\,p,b\right)=\sigma_{v}^{R*}\left(g\,|\,p,b\right)=\frac{\omega_{d}^{t}}{\left(q^{R}\left(0\right)+q^{R}\left(v\right)\right)\delta_{o}\left(b\,|\,d,p\right)\left(\omega_{d}^{t}-\omega_{d}^{g}\right)},\label{eq:eq3ProbG}
\end{equation}
 and trust with probability $\sigma_{o}^{R*}\left(t\,|\,p,b\right)=\sigma_{v}^{R*}\left(t\,|\,p,b\right)=1-\sigma_{o}^{R*}\left(g\,|\,p,b\right).$
Deceptive $S$ persist with reduced probability 
\begin{equation}
\sigma_{d}^{S*}\left(p\right)=\frac{1}{q^{S}\left(d\right)}\left(\frac{\tilde{U}_{o}^{R}\left(l,p,t\right)}{\tilde{U}_{o}^{R}\left(l,p,t\right)-\tilde{U}_{o}^{R}\left(d,p,t\right)}\right).\label{eq:eq3decep}
\end{equation}

The\emph{ status quo} and \emph{resistant attacker }equilibria are
poor results because infection of devices is not deterred at all.
The focus of Section \ref{sec:Mechanism-Design} will be to shift
the PBNE to the other equilibrium regions, in which infection of devices
is deterred to some degree.
\begin{table}
\caption{\label{tab:eqRegion}Equilibrium Regions of the PSG for PDoS}

\centering{}%
\begin{tabular}{c|c|c|}
 & $q^{S}(d)<TD_{v}^{R}(\bullet)$  & $q^{S}(d)>TD_{v}^{R}(\bullet)$\tabularnewline
\hline 
\multirow{2}{*}{$BP_{d}^{S}(\bullet\,|\,t,g,g)<0,$} & \multicolumn{2}{c|}{\uline{Vulnerable Attacker}}\tabularnewline
 & \multicolumn{2}{c|}{$\sigma^{S*}(p)<1$}\tabularnewline
\multirow{2}{*}{$BP_{d}^{S}(\bullet\,|\,t,g,f)<0$} & \multicolumn{2}{c|}{$0<\sigma_{o}^{R*}(t\,|\,p,b),\,\sigma_{o}^{R*}(g\,|\,p,b)<1$}\tabularnewline
 & \multicolumn{2}{c|}{$0<\sigma_{v}^{R*}(t\,|\,p,b),\,\sigma_{v}^{R*}(g\,|\,p,b)<1$}\tabularnewline
\hline 
\multirow{2}{*}{$BP_{d}^{S}(\bullet\,|\,t,g,g)>0,$} &  & \uline{Active Deterrence}\tabularnewline
 &  & $\sigma^{S*}(p)<1$\tabularnewline
\multirow{2}{*}{$BP_{d}^{S}(\bullet\,|\,t,g,f)<0$} & \uline{Status Quo}  & $\sigma_{o}^{R*}(g\,|\,p,b)=1$\tabularnewline
 & $\sigma^{S*}(p)=1$  & $0<\sigma_{v}^{R*}(g\,|\,p,b),$\tabularnewline
 & $\sigma_{o}^{R*}(g\,|\,p,b)=1$  & $\sigma_{v}^{R*}(f\,|\,p,b)<1$\tabularnewline
\cline{1-1} \cline{3-3} 
\multirow{2}{*}{$BP_{d}^{S}(\bullet\,|\,t,g,g)>0,$ } & $\sigma_{v}^{R*}(g\,|\,p,b)=1$  & \uline{Resistant Attacker}\tabularnewline
 &  & $\sigma^{S*}(p)=1$\tabularnewline
\multirow{2}{*}{$BP_{d}^{S}(\bullet\,|\,t,g,f)>0$} &  & $\sigma_{o}^{R*}(g\,|\,p,b)=1$\tabularnewline
 &  & $\sigma_{v}^{R*}(f\,|\,p,b)=1$\tabularnewline
\hline 
\end{tabular}
\end{table}

\section{Mechanism Design\label{sec:Mechanism-Design}}

The equilibrium results are delineated by the quantities $q^{S},$
$TD_{v}^{R}(U_{v}^{R},\delta_{v})$ and $BP_{d}^{S}(\omega_{d},q^{R},\delta).$
These quantities are functions of the parameters $q^{S},$ $q^{R},$
$\delta_{o},$ $\delta_{v},$ $\omega_{d},$ and $U_{v}^{R}.$ Mechanism
design manipulates these parameters in order to obtain a desired equilibrium.
We discuss two possible mechanisms.
\begin{figure}
\centering{}\includegraphics[width=0.48\columnwidth]{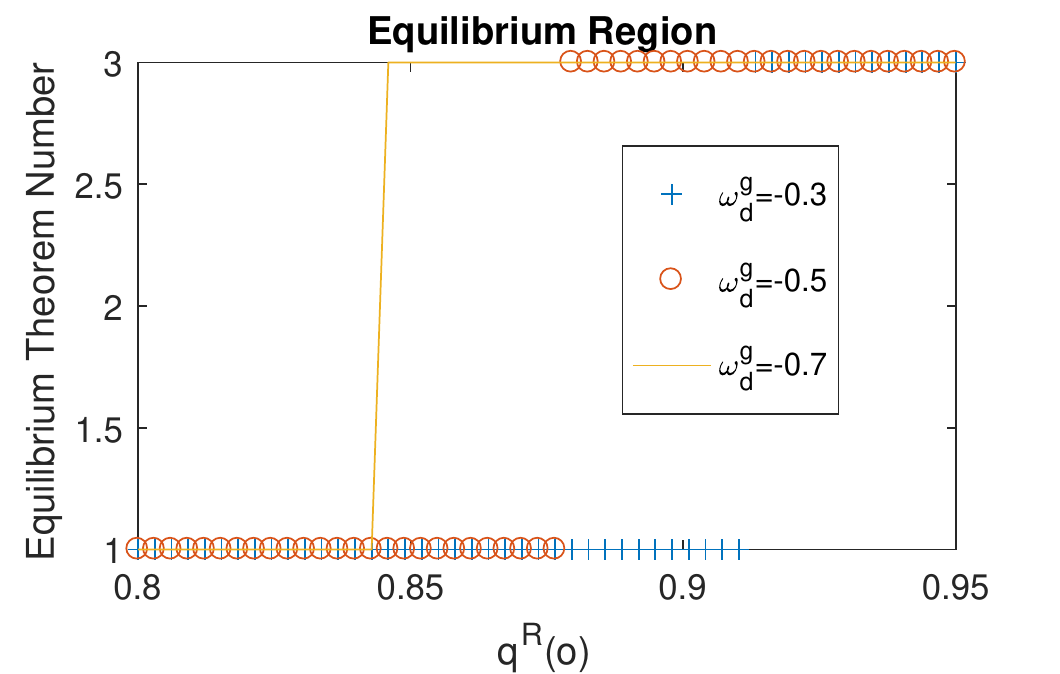}\hfill{}\includegraphics[width=0.48\columnwidth]{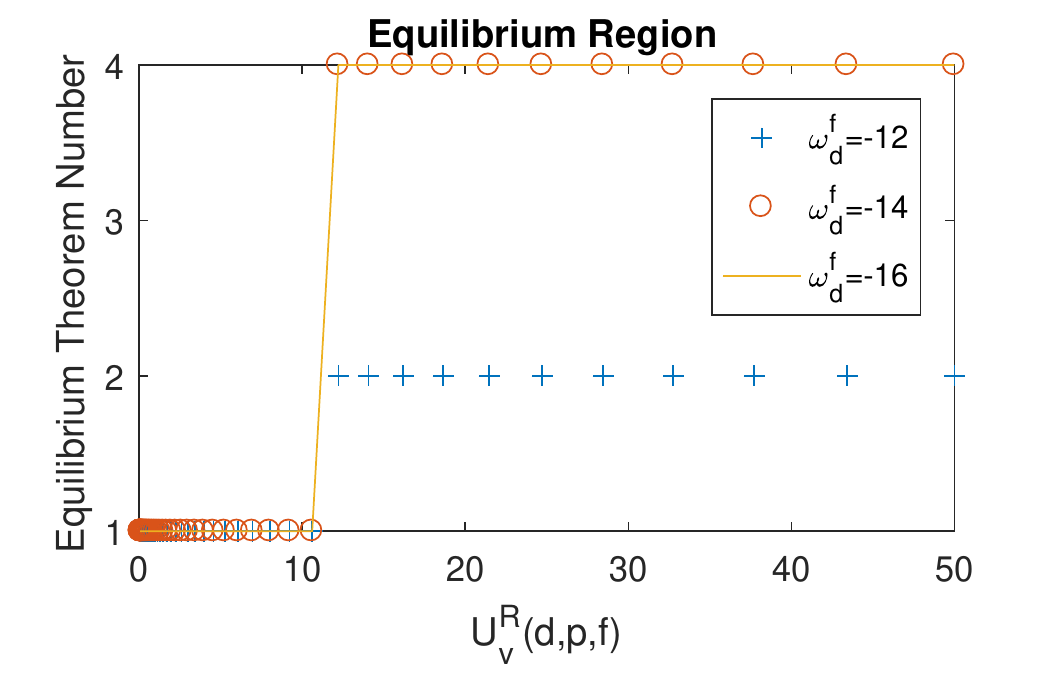}\caption{\label{fig:legalEq}Equilibrium transitions for (a) legal and (b)
active defense mechanisms. The equilibrium numbers signify: 1-\emph{status
quo}, 2-\emph{resistant attacker}, 3-\emph{vulnerable attacker}, 4-\emph{active
deterrence}.}
\end{figure}
\begin{figure}
\centering{}\includegraphics[width=0.48\columnwidth]{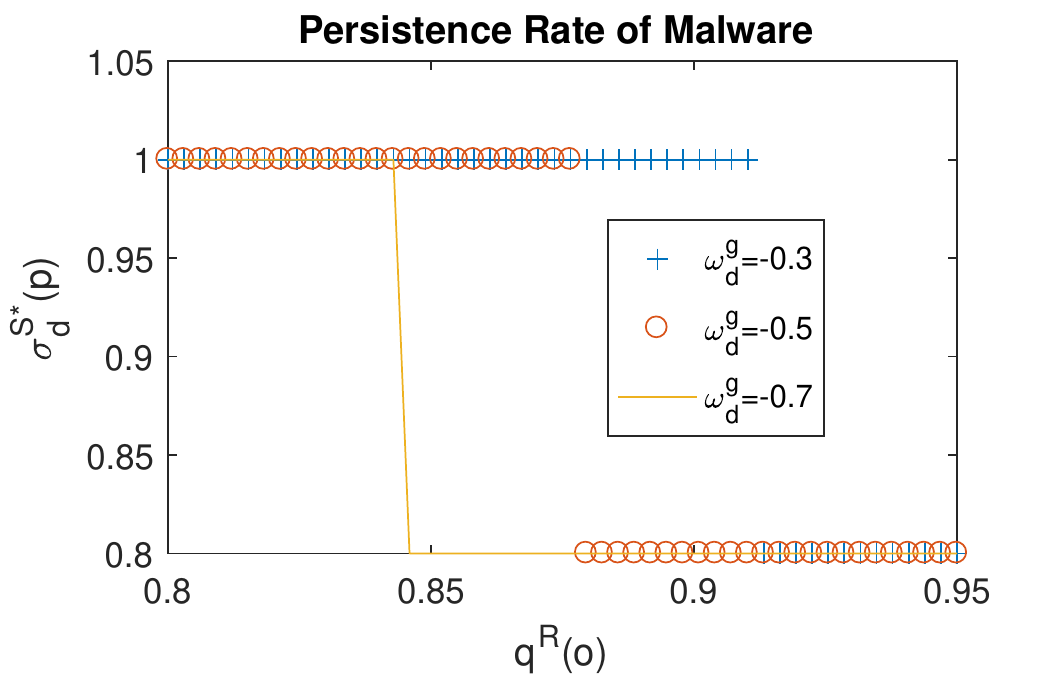}\hfill{}\includegraphics[width=0.48\columnwidth]{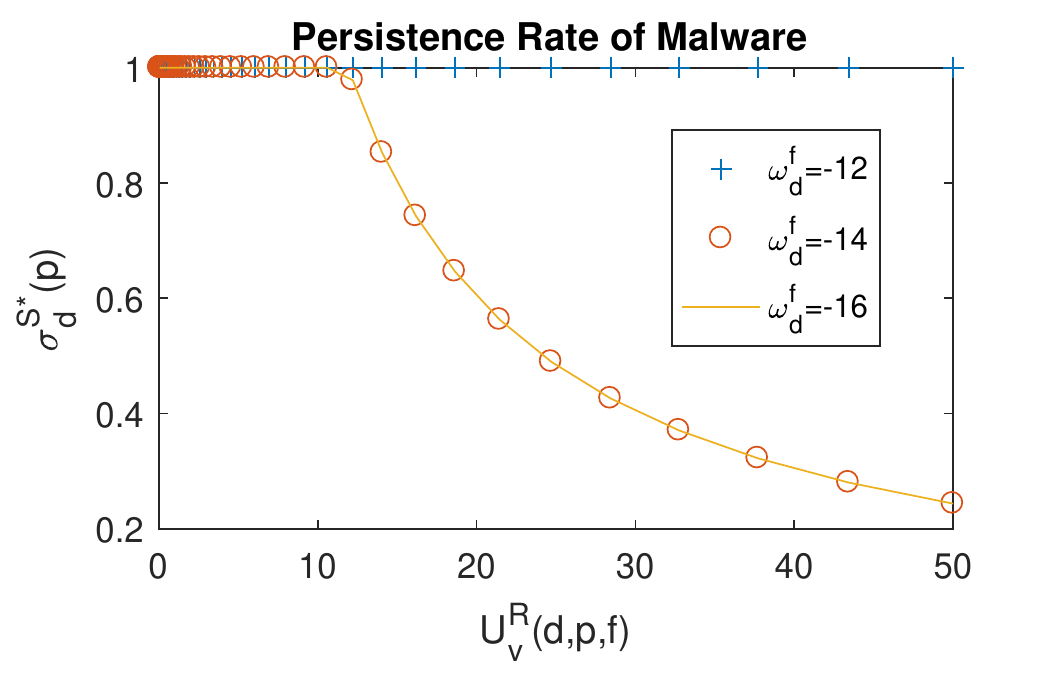}
\caption{\label{fig:legalPartic}Malware persistence rate for (a) legal and
(b) active defense mechanisms. }
\end{figure}
\begin{figure}
\centering{}\includegraphics[width=0.48\columnwidth]{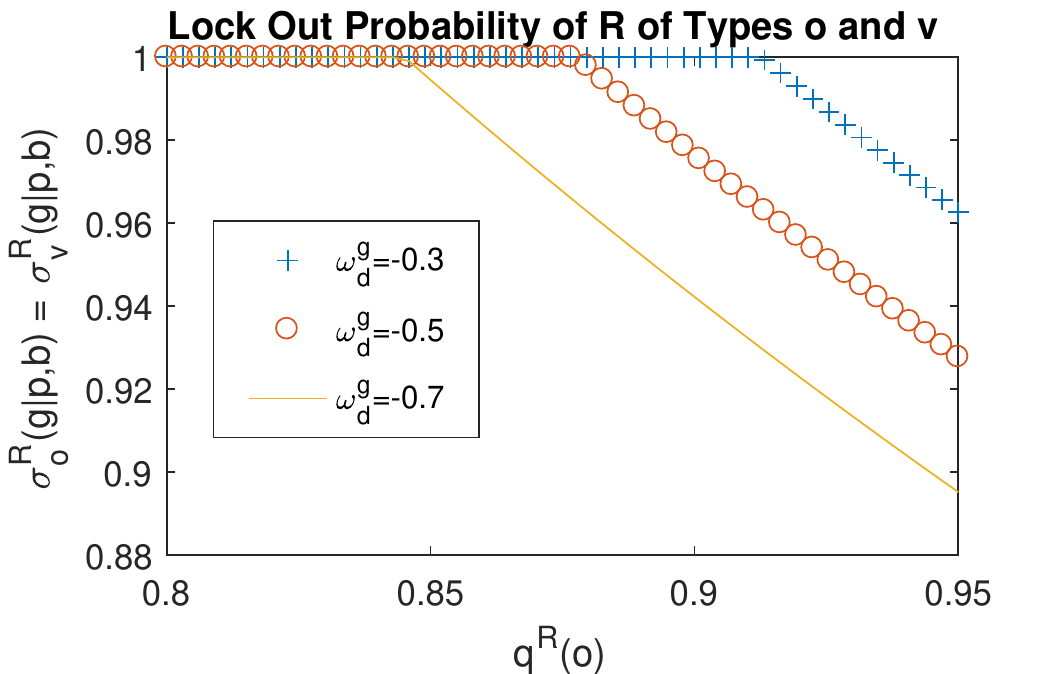}\hfill{}\includegraphics[width=0.48\columnwidth]{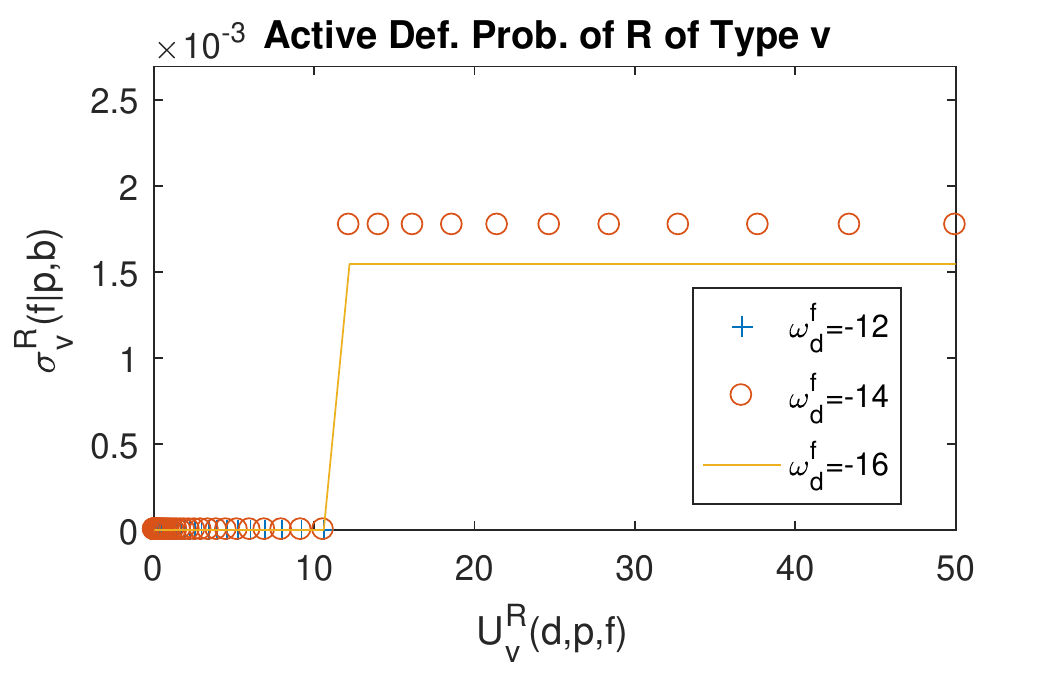}
\caption{\label{fig:legalIgnores}Probabilities of opposing malicious $S.$
Plot (a): probability that $R$ lock down with the legal mechanism.
Plot (b): probability that $R$ use active defense.}
\end{figure}

\subsection{Legislating Basic Security}

Malware which infects IoT devices is successful because many IoT devices
are poorly secured. Therefore, one mechanism design idea is to legally
require better authentication methods, in order to decrease $q^{R}(k)$
and increase $q^{R}(o).$

The left-hand sides of Figs. \ref{fig:legalEq}-\ref{fig:legalIgnores}
depict the results. Figure \ref{fig:legalEq}(a) shows that decreasing
$q^{R}(k)$ and increasing $q^{R}(o)$ moves the game from the \emph{status
quo} equilibrium to the \emph{vulnerable attacker }equilibrium. But
Fig. \ref{fig:legalPartic}(a) shows that this only causes a fixed
decrease in $\sigma_{d}^{S*}(p),$ regardless of the amount of decrease
in $q^{R}(k).$ The reason, as shown in Fig. \ref{fig:legalIgnores}(a),
is that as $q^{R}(o)$ increases, it is incentive-compatible for receivers
to lock down with progressively lower probability $\sigma_{y}^{R*}(g\,|\,p,b),$
$y\in\{o,v\}.$ Rather than forcing malicious $S$ to not persist,
increasing $q^{R}(o)$ only decreases the incentive for receivers
to lock down under suspicious login attempts.

\subsection{Incentivizing Active Defense }

One reason for the proliferation of IoT malware is that most devices
which are secure (\emph{i.e.}, $R$ of type $y=o$) do not take any
actions against malicious login attempts except to lock down (\emph{i.e.},
to play $a=g$). But there is almost no cost to malware scanners for
making a large number of login attempts under which devices simply
lock down. There is a lack of economic pressure which would force
$\sigma_{d}^{S*}(p)<1,$ unless $q^{R}(0)\approx1.$

This is the motivation for using active defense such as reporting
the activity to an ISP or recording the attempts in order to gather
information about the attacker. The right hand sides of Figs. \ref{fig:legalEq}-\ref{fig:legalIgnores}
show the effects of providing an incentive $U_{v}^{R}(d,p,f)$ for
active defense. This incentive moves the game from the \emph{status
quo} equilibrium to either the \emph{resistant attacker} equilibrium
or the \emph{vulnerable attacker }equilibrium, depending on whether
$BP_{d}^{S}(\bullet\,|\,t,g,f)$ is positive (Fig. \ref{fig:legalEq}(b)).
In the \emph{vulnerable attacker} equilibrium, the persistence rate
of malicious $S$ is decreased (Fig. \ref{fig:legalPartic}(b)). Finally,
Fig. \ref{fig:legalIgnores}(b) shows that only a small amount of
active defense $\sigma_{v}^{R*}(f\,|\,p,b)$ is necessary, particularly
for high values of\footnote{In Fig. \ref{fig:legalIgnores}(b), $\sigma_{v}^{R*}(f\,|\,p,b)=1$
for $\omega_{d}^{f}=-12.$ } $\omega_{d}^{f}.$

\section{Discussion of Results\label{sec:Discussion-of-Results}}

The first result is that \emph{the defender can bound the activity
level of the botnet}. Recall that\emph{ }the \emph{vulnerable attacker}
and \emph{active deterrence} equilibria force $\sigma_{d}^{S*}(p)<1.$
That is, they decrease the persistence rate of the malware scanner.
But another interpretation is possible. In Eq. (\ref{eq:eq3decep})
and Eq. (\ref{eq:eq4decep}), the product $\sigma_{d}^{S*}\left(p\right)q^{S}\left(d\right)$
is bounded. This product can be understood as the \emph{total activity}
of botnet scanners: a combination of prior probability of malicious
senders and the effort that malicious senders exert\footnote{A natural interpretation in an evolutionary game framework would be
that $\sigma_{d}^{S*}(p)=1,$ and $q^{S}(d)$ decreases when the total
activity is bounded. In other words, malicious senders continue recruiting,
but some malicious senders drop out since not all of them are supported
in equilibrium.}. Bensoussan \emph{et} \emph{al}. note that the operators of the Confiker
botnet of 2008-2009 were forced to limit its activity \cite{bensoussan2010game,higgins2010confiker}.
High activity levels would have attracted too much attention. The
authors of \cite{bensoussan2010game} confirm this result analytically,
using a dynamic game based on an SIS infection model. Interestingly,
our result agrees with \cite{bensoussan2010game}, but using a different
framework. 

Secondly, we\emph{ compare the effects of legal and economic mechanisms
to deter recruitment for PDoS}. Figures \ref{fig:legalEq}(a)-\ref{fig:legalIgnores}(a)
showed that $\sigma_{d}^{S*}(p)$ can only be reduced by a fixed factor
by mandating security for more and more devices. In this example,
we found that strategic behavior worked against legal requirements.
By comparison, Figs. \ref{fig:legalEq}(b)-\ref{fig:legalIgnores}(b)
showed that $\sigma_{d}^{S*}(p)$ can be driven arbitrarily low by
providing an economic incentive $U_{v}^{R}(d,p,f)$ to use active
defense. 

Future work can evaluate technical aspects of mechanism design such
as improving malware detection quality. This would involve a non-trivial
trade-off between a high true-positive rate and a low false-positive
rate. Note that the model of Poisson signaling games is not restricted
PDoS attacks. PSG apply to any scenario in which one sender communicates
a possibly malicious or misleading message to an unknown number of
receivers. In the IoT, the model could capture the communication of
a roadside location-based service to a set of autonomous vehicles,
or spoofing of a GPS signal used by multiple ships with automatic
navigation control, for example. Online, the model could apply to
deceptive opinion spam in product reviews. In interpersonal interactions,
PSG could apply to advertising or political messaging.  

\appendix

\section{Simplification of Sender Expected Utility \label{apdx:simpSeUtil}}

Each each component of $c$ is distributed according to a Poisson
r.v. The components are independent, so $\mathbb{P}\{c\,|\,\sigma^{R},x,m\}=\underset{a\in A}{\prod}\mathbb{P}\{c_{a}\,|\,\sigma^{R},x,m\}.$
Recall that $S$ receives zero utility when he plays $m=w.$ So we
can choose $m=p:$ 
\[
\bar{U}_{x}^{S}(\sigma_{x}^{S},\sigma^{R})=\sigma_{x}^{S}\left(p\right)\underset{c\in\mathbb{Z}\left(A\right)}{\sum}\,\underset{a\in A}{\prod}\mathbb{P}\left\{ c_{a}\,|\,\sigma^{R},x,p\right\} \left(\omega_{x}^{t}c_{t}+\omega_{x}^{g}c_{g}+\omega_{x}^{f}c_{f}\right).
\]
Some of the probability terms can be summed over their support. We
are left with 
\begin{equation}
\bar{U}_{x}^{S}(\sigma_{x}^{S},\sigma^{R})=\sigma_{x}^{S}\left(p\right)\sum_{a\in A}\omega_{x}^{a}\sum_{c_{a}\in\mathbb{Z}_{+}}c_{a}\mathbb{P}\left\{ c_{a}\,|\,\sigma^{R},x,p\right\} .\label{eq:intermedExpUtilS}
\end{equation}
The last summation is the expected value of $c_{a},$ which is $\lambda_{a}.$
This yields Eq. (\ref{eq:uBarStract}).

\section{Proof of Theorem \ref{thm:eqInRegions}\label{apdx:Thm1}}

The proofs for the \emph{status quo} and \emph{resistant attacker}
equilibria are similar to the proof for Lemma \ref{lem:eqAfterAssump}.
The \emph{vulnerable attacker} equilibrium is a partially-separating
PBNE. Strategies $\sigma_{o}^{R*}(g\,|\,p,b)$ and $\sigma_{v}^{R*}(g\,|\,p,b)$
which satisfy Eq. (\ref{eq:eq3ProbG}) make malicious senders exactly
indifferent between $m=p$ and $m=w.$ Thus, they can play the mixed-strategy
in Eq. (\ref{eq:eq3decep}), which makes strong and active receivers
exactly indifferent between $a=g$ and $a=t.$ The proof of the \emph{vulnerable
attacker} equilibrium follows a similar logic.

\bibliographystyle{plain}
\bibliography{PDoSjBib}

\end{document}